\documentclass[12pt]{article}%
\usepackage{ amsmath, graphics, rotating}%

\begin{document}
\large 
\begin{center}
{\bf Einstein-Aether Theory With and Without Einstein}
\end{center}

\begin{center}
{Boris Hikin}
\small

E-mail:  blhikin@gmail.com

Tel: 310-922-4752 or 310-826-0209, USA \end{center}
\vskip 1em
\vskip 0.5em
\small

The exact static spherically symmetric solutions for pure-aether theory and Einstein-aether theory are presented. 
It is shown that both theories can deliver the Schwarzschild  metric, but only the Einstein-aether theory
contains solutions with "almost-Schwarzschild" metrics that satisfy Einstein's experiments.
Two specific solutions are of special interest: one in pure-aether theory that derives the attractive nature
of gravitation as a result of Minskowski signature of the metric, and one - the Jacobson solution-  of Einstein-aether
theory with "almost-Schwarzschild" metric and non-zero Ricci tensor.

\baselineskip 1pc
\large
\vskip 3em
{\underline {Introduction}}
\vskip 1em

Einstein-aether theory was proposed by Jacobson and Mattingly almost 10 years ago \cite{i1} (see latest review
\cite{i2}) as a correction (extension) of Einstein's GR. It postulates that gravitation, in addition to
a curved space, is characterized by a unit vector $G_i$ ($G_iG_jg^{ij}=1$), 
which together with metric tensor $g_{ij}$ constitute the dynamic 
variables of the Einstein-aether theory. The equations of motion for $g_{ij}$ and for $G_i$ are 
derived by variational method from a Lagrangian that is a function of the metric tensor
$g_{ij}$ and the vector field $G_i$.

With the requirement that the equations of motion are of the second order, the most general form of Lagrangian
can be written as \cite{i1}:
\begin{eqnarray}
\label{f5_1}
&&S=\int\nolimits L(g_{ij}, G_i) {\sqrt{-g}} d^4x \nonumber \\
&&L=c_0R+c_1G_{i;j}G^{i;j}+c_3G_{i;j}G^{j;i}+c_2(G^i_{;i})^2+c_4G^{k;i}G_iG_{k;j}G^j\nonumber\\
&&\quad +T(G_iG_jg^{ij}-1)
\end{eqnarray}

where all c's are constants, T is the Lagrange multiplier, and R is Ricci scalar.

The main question that Jacobson and Mattingly raised was: what should the value of the constants "c"s be in order for
the Einstein-aether theory to yield the same results as predicted by GR? In particular, in case of spherical symmetry the 
Einstein-aether theory must yield - in the first order of approximation - the Schwarzschild's metric.

The Lagrangian of the aether theory, eq. (\ref{f5_1}), actually represens three separate theories: 

	a) General Relativity $c_0 \neq 0$ all other c's are zero ($c_1=c_2=c_3=c_4=0$)

	b) Einstein-aether theory $c_0 \neq 0$ and at least one of others c's (typically $c_1$) is not zero

	c) Pure-aether theory with $c_0=0$ and at least one of others c's is not zero

One question that has not been addressed previously is the validity of pure-aether theory.
In other words, do we need Einstein's term ($c_0R$) in the Lagrangian? One might argue that the 
Einstein's term ($c_0R$) is not needed and the pure-aether theory is sufficient to explain
the experimental results - at least as far as the solar system is concerned.

In this paper we would like to investigate
this issue based on spherically symmetric solutions and we will show that on one hand the Einstein term is not needed - 
if one wants to get Schwarzschild's metric solution - and on the other hand, the Einstein term leads to unique solutions
that are not present in pure-aether theory.

\vskip 3em
{\underline {Gravitation. Static, spherically symmetric solution.}}
\vskip 1em
In a case of spherical symmetry the last term of Lagrangian (the $c_4$ term) can be expressed thru the
other terms (see \cite{i3}) and thus can be dropped out.

The remaining four terms can be written in the following manner:

\begin{eqnarray}
\label{f5_2}
&&S=\int\nolimits L(g_{ij}, G_i) {\sqrt{-g}} d^4x\nonumber \\
&&L=-[\lambda_0R+\lambda_1g^{ik}g^{jl}(\partial_jG_i-\partial_iG_j)(\partial_lG_k-\partial_kG_l) \nonumber\\
&&\quad +\lambda_2R_{ij}G^iG^j+\lambda_3(G^k_{;k})^2 +T(G_iG_jg^{ij}-1)]
\end{eqnarray}

where the $\lambda$s are new constants that are linear combinations of c-constants.

The equations of motion for the metric $g_{ij}$ and for the vector field $G_i$ have been obtained previously \cite {i3}.
However, it is quite a laborious task to derive the exact solution using these equations.
Fortunately, in case of spherical symmetry, there is another approach that yields the desired result with almost
no hard labor involved.

In a case of static (time independent) spherically symmetric solution the aether theory is governed by 5 functions:
$g_{00}$ (or $g_{tt}$), $g_{11}$ (or $g_{rr}$), $G_0$ (or $Q_t$), $G_1$ (or $Q_r$) and T, 
which are all functions of radius only. We assume that $g_{22}=-r^2$

By using the constraint $G_iG^i=1$, one can eliminate two of these five functions - $g_{00}$ and T. Our next step is
to write the Lagrangian thru the three remaining functions ($G_1$, $g_{11}$ and $G_0$). 
And then find the equations for these functions by 
means of variation of the Lagrangian with respect to these functions.

The problem is simplified if we choose new variables in this manner:
\begin{eqnarray}
\label{f5_3}
&&\hat g=-g_{11}g_{00}\quad instead\,\,of\,\,g_{11}\nonumber\\
&&\bar  G_1=G_1 \frac {\sqrt{g_{00}}} {\sqrt{-g_{11}}}\quad instead\,\,of\,\, G_1\nonumber\\
&&x=\frac{1}{r}; \quad \frac {d}{dx}=()'
\end{eqnarray}

In these new variables $g_{00}$ can be written as :
\begin{eqnarray}
\label{f5_4}
&&g_{00}=(G_0)^2-(\bar  G_1)^2
\end{eqnarray}
and the Lagrangian (eq. \ref{f5_2}) has this (amazingly) simple form (for details see appendix A):

\begin{eqnarray}
\label{f5_5}
&&S=-\int\nolimits d^4v \sqrt{-g} L =\int\nolimits dtd \Omega S_r, \quad where\nonumber\\
&&S_r=-\int\nolimits r^2dr \sqrt{\hat g} L(g_{ij}, G_i)=\int\nolimits \frac{dx}{x^4} \sqrt{\hat g} L=\\
&&\int\nolimits dx\{\lambda_0[\frac{2}{\sqrt{\hat g}} (\frac{G^2_0}{x})'-\frac{2\sqrt{\hat g}}{x^2}]
-\frac{2\lambda_1(G_0')^2}{\sqrt{\hat g} }- \frac{2\bar \lambda_2}{\sqrt{\hat g}}(\frac {{\bar  G_1}^2}{x})'
+\frac{\lambda_3}{\sqrt{\hat g}}[x^2(\frac{\bar  G_1}{x^2})']^2 \} \nonumber
\end{eqnarray}
where $\bar \lambda_2=\lambda_2+\lambda_0$

We now consider the solution of this problem with the following boundary conditions:

	a) $G_i$ at infinity has only the time component:
\begin{eqnarray}
\label{f5_6a}
&&{G_i}_{|x=0\,\, or\,\, r=\infty }=(1,0,0,0)\quad or\nonumber\\
&&G_{0|x=0}=1; \quad G_{1|x=0}=\bar G_{1|x=0}=0
\end{eqnarray}

	b) metric at infinity corresponds to a flat space:
\begin{eqnarray}
\label{f5_6b}
&&{g_{ij}}_{|x=0\,\, or\,\, r=\infty }=diag(1,-1,-r^2,-r^2 sin^2 \theta)
\end{eqnarray}

\vskip 1em
The requirement that metric $g_{ij}$ satisfies two Einstein experiments (bending of light and precession of Mercury)
set these conditions on $\hat g$ (bending of light) and $g_{00}$ (precession of Mercury)
 as functions of x (1/r) with $x\approx 0$ (r$ \rightarrow \infty$) 
\cite{i4}:

\begin{eqnarray}
\label{f5_7}
&&\hat g=1+\bar cx^2+... (\bar c=constant) \nonumber\\
&&g_{00}=1+c_1x+c_2x^3+... (c_1\,and \, c_2=constant)
\end{eqnarray}
In other words, $g_{00}$ has no quadratic terms in seria by x and $\hat g$ has no linear terms.
\vskip 1em
\underline {The case of General Relativity}

Let us - mostly to demonstrate the simplicity of this approach and as a sanity check - 
first consider the case of GR ($\lambda_1=\lambda_2=\lambda_3=0$).

Variation of eq. (\ref{f5_5}) with respect to $G_0$ and $\hat g$ yields:

\begin{eqnarray}
\label{f5_8}
&&\frac{\delta S_r}{\delta G_0}=0 \quad ->\quad {\hat g}'=0 \quad or \quad {\hat g}=1\nonumber\\
&&\frac{\delta S_r}{\delta \hat g}=0 \quad ->\quad (\frac{G^2_0}{x})'+\frac{1}{x^2}=0\quad or \quad g_{00}=G^2_0=1+xC_0
\end{eqnarray}

The above solutions are exactly the expression of the Schwarzschild metric.
It is worth pointing out that the sign of the constant $C_0$ in GR is not set by the theory, but is taken
as a seperate postulate that gravitation is only attractive. The GR theory by itself does allow both attraction ($C_0<0$)
and repulsion ($C_0>0$).

\vskip 1em
\underline {The case of Pure-aether theory}

Let us now consider the case of pure aether theory ($\lambda_0=0$). 
Variation of eq. (\ref{f5_5}) with respect to $G_0$ $\bar G_1$, and $\hat g$ yields the following equations:

\begin{eqnarray}
\label{f5_9}
&&a) \quad \frac{\delta S_r}{\delta G_0}=0 \quad \rightarrow \quad (\frac{G'_0}{\sqrt{\hat g}})'=0\nonumber\\
&&b) \quad \frac{\delta S_r}{\delta \bar G_1}=0 \quad \rightarrow \quad 2\lambda_2(\frac {1}{\sqrt{\hat g}})'
\frac{\bar G_1}{x}-\frac{\lambda_3}{x^2}[\frac{x^4}{\sqrt{\hat g}}(\frac{\bar G_1}{x^2})']'=0\\
&&c) \quad \frac{\delta S_r}{\delta \hat g}=0 \quad \rightarrow \quad \lambda_1 (G'_0)^2+\lambda_2(\frac{\bar G^2_1}{x})'
-\frac{\lambda_3}{2}[x^2 (\frac{\bar G_1}{x^2})']^2 =0\nonumber
\end{eqnarray}

The first equation (eq. \ref{f5_9}a) can be integrated and with the third equation (eq. \ref{f5_9}c)
it can be used to find the function $\hat g$.
\begin{eqnarray}
\label{f5_10}
&&a) \quad G'_0=C_0\sqrt{\hat g} \quad C_0=constant\nonumber\\
&&b) \quad \lambda_3[\bar G''_1-\frac{2\bar G_1}{x^2}]
=[\frac{\lambda_3}{2}(\bar G'_1-\frac{2\bar G_1}{x})- \lambda_2 \frac{\bar G_1}{x} ]  \frac{{\hat g}'}{\hat g}\\
&&c) \quad \hat g=-\frac{ \lambda_2}{C^2_0\lambda_1}(\frac{\bar G^2_1}{x})'
+\frac{\lambda_3}{2C^2_0\lambda_1}[x^2 (\frac{\bar G_1}{x^2})']^2 \nonumber
\end{eqnarray}

It is not difficult to see that if $\lambda_3$ is not zero
 the system of eqs. (\ref{f5_10}) has no solutions
that satisfy the conditions (\ref{f5_7}).
Indeed, if $\hat g \approx 1+x^2\bar c$ then $\hat g' \approx x \bar c$ and the rhs of eq. (\ref{f5_10}b) is about constant or zero.
In order for the lhs of (\ref{f5_10}b) to be regular, $\bar G_1$ is about $x^2$. 
In this case the $\hat g$ per eq. (\ref{f5_10}c)
should be zero at x=0, which contradicts to the requirements (\ref{f5_7}).

With $\lambda_3=0$, the system of equations (\ref{f5_10}) can be easily integrated to give this result:
\begin{eqnarray}
\label{f5_11}
&&\hat g=1; \quad \bar G_0=1+C_0x; \quad \bar G^2_1=C_1x-\frac{\lambda_1}{\lambda_2}C^2_0x^2\nonumber\\
&&and \quad g_{00}=1+(2C_0-C_1)x+(1+\frac{\lambda_1}{\lambda_2})C^2_0x^2
\end{eqnarray}

In the above expressions $C_0$ and $C_1$ are constant and $C_1>0$.

In order to satisfy the requirements of Einstein's experiments, eq. (\ref{f5_7}), the quadratic term must be set to zero,
which can be achieved if $C_0=0$ or $\lambda_1+\lambda_2=0$. In both cases the metric is the Schwarzschild one.

If one sets $C_0=0$ the time component of vector field is one ($Q_0=1$) and the radius component $\bar G_1$ is inverse
to square root of the radius r :
\begin{eqnarray}
\label{f5_12}
&&G_0=1; \,\,g_{00}=-\frac{1}{g_{11}}=1-\frac{C_1}{r};\,\, G_1= \sqrt{\frac{\bar C_1}{ r(1-C_1/r )^2} }
\end{eqnarray}
In addition to Schwarzschild metric the pure-aether  theory (with the condition $\lambda_3=0$) 
delivers the requirement that gravitation must be attractive,
which is the consequence of Minkowski signature of the metric tensor.

\vskip 2em
\underline {Einstein-aether theory ($\lambda_0 \neq 0$)}
\vskip 1em
We now can consider the Einstein-aether theory, or the case when $\lambda_0$ is not zero.
The presence of Einstein term ($\lambda_0R$) in Lagrangian significantly changes the number of
 possible solutions and the choice of $\lambda$ parameters.

The variation of action integral $S_r$, eq. (\ref{f5_5}), leads to this set of equations:

\begin{eqnarray}
\label{f5_13}
&&a) \quad \frac{\delta S_r}{\delta G_0}=0 \quad \rightarrow 
\quad \lambda_1(\frac{G'_0}{\sqrt{\hat g} } )'-\lambda_0 \frac{G_0}{x}  (\frac{1}{\sqrt{\hat g}})'=0\nonumber\\
&&b) \quad \frac{\delta S_r}{\delta \bar G_1}=0 \quad \rightarrow \quad 2\bar \lambda_2(\frac {1}{\sqrt{\hat g}})'
\frac{\bar G_1}{x}-\frac{\lambda_3}{x^2}[\frac{x^4}{\sqrt{\hat g}}(\frac{\bar G_1}{x^2})']'=0\nonumber\\
&&c) \quad \frac{\delta S_r}{\delta \hat g}=0 \quad \rightarrow \\
&&\frac{\lambda_0}{\sqrt{\hat g^3}} (\frac {G^2_0}{x})'+\frac{\lambda_0}{x^2\sqrt{\hat g}} 
-\frac{\lambda_1}{\sqrt{\hat g^3}} (G'_0)^2
-\frac{\bar \lambda_2}{\sqrt{\hat g^3}} (\frac{\bar G^2_1}{x})'
+\frac{\lambda_3}{2 \sqrt{\hat g^3}}
 [x^2(\frac{\bar G_1}{x^2})']^2 =0\nonumber
\end{eqnarray}

From the last equation (\ref{f5_13}c) one can express $\hat g$ as a function of $G_0$ and $\bar G_1$:
\begin{eqnarray}
\label{f5_14}
&&\hat g=x^2\{-(\frac {G^2_0}{x})'+\frac{\lambda_1}{\lambda_0} (G'_0)^2  
+\frac{\bar \lambda_2}{\lambda_0} (\frac{G^2_1}{x})'-\frac{\lambda_3}{2\lambda_0} [x^2(\frac{\bar G_1}{x^2})']^2 \}
\end{eqnarray}

Because of the $x^2$ factor in front of the figure braket it is not difficult to see that for any $\lambda$s
 the expression for $\hat g$ is always regular (no singularities).

Let us note that if $\lambda_3$ is not zero
from the equation (\ref{f5_13}b) follows that for $x \rightarrow 0$
$\bar G_1 \approx x^2$ and the $\lambda_2$, $\lambda_3$ terms of eq. (\ref{f5_14}) are about $x^4$.

This means that if we are interested in the behavior of $\hat g$ near $x=0$ the
$\lambda_2$, $\lambda_3$ terms of eq. (\ref{f5_14}) could be dropped out. The remaining expression for $\hat g$
( function of $G_0$ only) always has a right behavior that satisfies the condition of eq. (\ref{f5_7}) at x=0.
Indeed for $\hat g$ we have:
\begin{eqnarray}
\label{f5_15}
&&\hat g=G^2_0- 2xG_0 G'_0-\lambda_1 x^2(G'_0)^2
\end{eqnarray}
If we write $G_0$ near zero as a series by x ,  $G_0=1+ax+bx^2+..$ , (a and b are constants) and substitute it in eq. (\ref{f5_15}) 
above we will get this approximation for $\hat g$:
\begin{eqnarray}
\label{f5_16}
&&\hat g=(1+ax+bx^2+...)^2-2x((1+ax+bx^2+...)(a+2bx+...)\nonumber\\
&&-\lambda_2x^4A^2-\lambda_3x^4A^2=1+(-a^2-2b-\lambda_1a^2)x^2+...
\end{eqnarray}
which as we see has no linear term and thus satisfies the condition (\ref{f5_7}).

In general the two equations, eq. (\ref{f5_13}), that describe variables $G_0$, $\bar G_1$ are coupled thru $\hat g$, which depends
on both functions. There are however four cases where the equations can be uncoupled and  there solutions can be
presented in analytical forms:

Case A: $\quad \lambda_3=0$ and $\bar G_1 \neq 0$

Case B: $\quad \bar G_1=0$ any $\lambda_1$, $\lambda_2$, $\lambda_3$

Case C: $\quad \bar \lambda_2=0$ (or $\lambda_2=-\lambda_0$)

Case D: $\quad \lambda_1=0$

{\underline {Case A}}
\vskip 1em

If $\lambda_3$ is zero (and $G_1 \neq 0$), the equation (\ref{f5_13}b) yields that $\hat g=1$ 
and from the equation (\ref{f5_13}a) follows that $G_0$ is a linear function of x ($G_0=1+C_0x$). 

The third equation (\ref{f5_14}) can be used to determine $\bar G_1$:
\begin{eqnarray}
\label{f5_17}
&&\hat g=G^2_0-2xG_0G'_0+\frac{\lambda_1}{\lambda_0}x^2(G'_0)^2+
\frac{\bar \lambda_2}{\lambda_0}x^2(\frac{\bar G^2_1}{x})' \quad or\nonumber\\
&&1=(1+C_0x)^2-2x(1+C_0x)C_0+\frac{\lambda_1}{\lambda_0}x^2C^2_0+
\frac{\bar \lambda_2}{\lambda_0}x^2(\frac{\bar G^2_1}{x})' \nonumber\\
&&\bar G^2_1=C_1x+C^2_0 x^2(\frac{\lambda_1-\lambda_0}{\bar \lambda_2})
\end{eqnarray}

 And for $g_{00}$ we get:
\begin{eqnarray}
\label{f5_17a}
&&g_{00}=G^2_0-\bar G^2_1 \quad or \nonumber\\
&&g_{00}=1+x(2C_0-C_1)+C^2_0x^2(1-\frac{\lambda_1-\lambda_0}{\bar \lambda_2})
\end{eqnarray}
This is practically (except for the value of the constants $\lambda$s)
the same result as for pure-aether theory that we derived above - eq.(\ref{f5_11}).

\vskip 1em
{\underline {Case B, ($G_1=0$) and C, ($\bar \lambda_2=0$)}}
\vskip 1em

In both of these cases the system of equations (\ref{f5_13}) can be solved analytically.

In the "case B" ($G_1=0$ and thus $\bar G_1=0$) the eq. (\ref{f5_13}b) is satisfied and in the remaining two equations
$\lambda_2$ and $\lambda_3$ terms could be dropped, leaving these equations for $G_0$ and $\hat g$:

\begin{eqnarray}
\label{f5_18}
&&\hat \lambda_1(\frac{G'_0}{\sqrt{\hat g} } )'- \frac{G_0}{x}  (\frac{1}{\sqrt{\hat g}})'=0
\quad or \quad G''_0=(\frac{G'_0}{2}-\frac{G_0}{2\hat \lambda_1x})\frac{\hat g'}{\hat g} \\
\label{f5_18a}
&&\quad \hat g=G^2_0- 2xG_0 G'_0+\hat \lambda_1 x^2(G'_0)^2 \quad 
where \quad \hat \lambda_1=\frac{\lambda_1}{\lambda_0}
\end{eqnarray}

In the "case C" ($\bar \lambda_2=0)$ the eq. (\ref{f5_13}b) has the solution $\bar G_1=C_1x^2$ ($C_1$ - constant), while
$G_0$ and $\hat g$ are defined by the same set of equations (\ref{f5_18}), (\ref{f5_18a}) as in "case B".

The equations (\ref{f5_18}), (\ref{f5_18a}) - although in slightly different form - had been obtained and investigated
by Jacobson in his 2006 paper \cite{i1}.
The equations can be integrated analytically to yield a result in a form $x=f(G_0)$ (for details see Appendix B):

\begin{eqnarray}
\label{f5_19}
&&C_0x=G_0[(G_0^{-\mu}-G_0^{\mu}], \,\, where \quad \mu=\sqrt{1-\frac{\lambda_1}{\lambda_0}}\nonumber\\
&&\hat g=\frac{4\mu^2 G^2_0}{[(1-\mu)G^{-\mu}_0-(1+\mu)G^{\mu}_0]^2}
\end{eqnarray}
where $C_0$ is a constant equivalent to Schwarzschild radius.

By direct calculation it is not difficult to show that for small x ($x \approx 0$) $\hat g$ has no linear terms 
($\hat g=1+ax^2+...$) and metric has no quadratic terms ($g_{00} \equiv G^2_0 \approx 1-C_0x+bx^3+...$) thus satisfying requirements
of the Einstein experiments, eq. (\ref{f5_7}), for any parameter $\mu$.

The behavior of $G_0$ vs. x outside $x=0$ (small distance r) 
significantly depends on a sign of $\lambda_1$ (we assume - as in GR - $\lambda_0>0$).

If $\lambda_1$=0 (the case of GR) the eqs.(\ref{f5_19}) and (\ref{f5_4}) yield:
\begin{eqnarray}
\label{f5_20}
&&\quad C_0x=1-G^2_0 \quad \hat g=1 \quad G_1=0\nonumber\\
&&g_{00}=G^2_0=1-C_0x \quad C_0=const
\end{eqnarray}
with the horizon point at $x=1/C_0$.

If $\lambda_1<0$, x as a function of $G_0$ monotonically increases as $G_0$ decreases from 1 to 0. This means that
$G_0$ as a function of x monotonically decreases from 1 to 0 as x changes from x=0 to $x=\infty$.

For the "case B"  the metric, defined as $g_{00}=G^2_0$, also decreases from 1 to 0 as x changes from 0 to infinity 
(the horizon point is $x=\infty$ or r=0). For small $G_0$ the second term in eq. (\ref{f5_19}) could be dropped giving
this expression for $G_0$ as a function of x:

\begin{eqnarray}
\label{f5_19a}
&&C_0x \approx G_0^{1-\sqrt{(1-\hat \lambda_1)}}\quad \nonumber\\
&&G_0 \approx (\frac{1}{C_0x})^{\frac{1}{\sqrt{(1-\hat \lambda_1)}-1}}=(\frac{r}{C_0})^{\frac{1}{\sqrt{(1-\hat \lambda_1)}-1}}
 \quad \hat \lambda_1 \equiv \frac{\lambda_1}{\lambda_0}<0
\end{eqnarray}

In the "case C", on the other hand, the metric has additional term: $g_{00}=G^2_0-(C_1)^2x^4$, which always - due to Minkowski
signature -leads to existence of a horizon point a some point x, the value of which depends on the value of the constant $C_1$.

If $\lambda_1$ positive ($1>\hat \lambda_1>0$) - analogues to Maxwell theory, x as a function of $G_0$ has a bell shape between
two points $G_0=1$ and $G_0=0$
with its maximum at some point in between. This means that $G_0(x)$ exists only from $x=0$ to a certain point -"dead point".
It can be explicitly illustrated for the case of $\hat \lambda_1=8/9$:
\begin{eqnarray}
\label{f5_21}
&&C_0x=G_0[(G_0^{-\frac13}-G_0^{\frac13}]\quad or \quad G_0=(\frac{1+\sqrt{1-4C_0x}}{2})^\frac32 
\end{eqnarray}
with $x=1/(4C_0)$ being a "dead point".

For the metric again we have two possibilities:

$\quad$ "Case B": $g_{00}=G^2_0$ and the metric exists up to a "dead point", which is not a horizon point, since $g_{00}$
is not zero.

$\quad$ "Case C": $g_{00}=G^2_0-(C_1)^2x^4$ and for sufficiently large $C_1$ metric ($g_{00}$) reaches zero - horizon point -
at some point before the "dead point". One can choose $C_1$ in such a way that at "dead point" ($x=1/4C_0$) 
the time component of the metric ($g_{00}$) become zero. That would represent the case when the "dead point" is the horizon.

\vskip 1em
{\underline {Case D, $\lambda_1=0$}}
\vskip 1em
We add this case mostly for the sake of completeness. The condition that $\lambda_1=0$ is probably non-physical,
due to the fact that equations of motion for the vector field $G_i$ become of the first - instead of second - order:
$aR^k_jG_k+b(G^k_{;k})_{,j}=0$ with a and b being constants. On the other hand we must
remember that the parameter $\lambda_1$ in our considerations is a combination of two parameters - see eq. (\ref{f5_1})
- $c_1$ and $c_4$, which would canceled each other only in the case of spherical symmetry.
\vskip 1em

If $\lambda_1=0$, the eq. (\ref{f5_13}a) yields $\hat g=1$ and the equation for $\bar G_1$ can be solved to yield 
$\bar G_1=C_1x^2$. Knowing $\hat g$ and $\bar G_1$, we can determine the function $G_0$ from eq. (\ref{f5_14}):
\begin{eqnarray}
\label{f5_21a}
&&1=x^2\{-(\frac {G^2_0}{x})'+\frac{\bar \lambda_2}{\lambda_0} 3C^2_1x^2\}\quad 
or \quad G^2_0=1+C_0x+\frac{\bar \lambda_2}{\lambda_0}C^2_1x^4\nonumber\\
&&g_{00}=1+C_0x+(\frac{\bar \lambda_2}{\lambda_0}-1)C^2_1x^4
\end{eqnarray}

\vskip 3em
{\underline {General Case, $\lambda 's \neq 0$}}
\vskip 1em
In the general case (both $G_1$ and $\lambda 's$ are not zero) the solutions have behavior somewhat in between 
"case C" and "case D".
For the small x (large distance r) $G_0$ linearly decreases, while $\bar G_1$ increases
(in absolute value) as $x^2$. As x moves toward large numbers
($r \rightarrow 0$) the $G_0$ starts deviate from Jacobson's solution while $\bar G_1$ deviates from $x^2$.

The same is true for the time component of metric $g_{00}$. In addition, if radial component of the vector field $\bar G_1$
is present (not zero), the metric has horizon point, which is due to the Minkowski signature of the metric.

\vskip 2em
{\underline {Discussion and Conclusion}}
\vskip 1em
A we saw above the general solution for aether theory is characterized by two parameters $C_0$ and $C_1$.
The first one, $C_0$, typically sets the linear dependence of $g_{00}$ as a function of x with x$\rightarrow 0$ 
(r $\rightarrow \infty$) and thus can be identified with
a Schwarzschild radius. A much more difficult question is the meaning of the other parameter $C_1$, 
which  defines the magnitude of radial dependence of aether vector $G_1$. 

\vskip 1em

We have shown that if one requires from the aether theory to get Schwarzschild solution for the metric, 
one can choose both
pure-aether theory (no Einstein $\lambda_0R$ term in Lagrangian) and Einstein-aether theory
(with $\lambda_0R$ term)  with $\lambda_3=0$ parameter.

It also must be pointed out that the Ricci tensor in both pure-aether and Einstein-aether
(with exception for the Jacobson solution) theories is always proportional to the constant $C_1$ - the radial
component ($G_1$ or $G_r$) of vector field $G_i$.

\vskip 1em

There are two particular solutions of the aether theory that deserve special attention.

The first  one is the solution of pure-aether or Einstein-aether theory with $\lambda_3=0$ and $C_0=0$:
\begin{eqnarray}
\label{f5_22}
G_0=1; \quad \bar G_1=\sqrt{C_1x} \,\, (C_1>0)  \quad \hat g=1 \quad and \quad g_{00}=1-C_1x.
\end{eqnarray}
The presence of "hard" matter does not change the time component of the aether field, but only adds the radial component.

In this solution the atractive nature of gravitation is derived from the aether theory and
is due to the Minkowski signature of the space metric.

The second one is the Jacobson's solution given by eq. (\ref{f5_19}):
\begin{eqnarray}
\label{f5_23}
&&\bar G_1=0\nonumber\\
&&C_0x=G_0[(G_0^{-\mu}-G_0^{\mu}], \,\, where \quad \mu=\sqrt{(1-\frac{\lambda_1}{\lambda_0})}>1\nonumber\\
&&\hat g=\frac{{4\mu}^2 G^2_0}{[(1-\mu)G^{-\mu}_0-(1+\mu)G^{\mu}_0]^2} \quad and \quad g_{00}=G^2_0
\end{eqnarray}

As r changes from infinity toward zero, $G_0$ declines from 1 to zero. Here we have that the hard matter
"replaces" the aether. This is opposite to the situation in Maxwell electrodynamics where the vector potential
increases toward the center of the charge. 

The Jacobson's metric, eq. (\ref{f5_23}), has no horizon (or to be more precise its horizon point is $r=0$)
and it has singularity ($g_{00}=0$) at $r=0$, which of course is an artifact of point-mass consideration.

It is also worthwhile to mention that the distance from the point- mass (r=0) to any point along radius is finite.

\begin{eqnarray}
\label{f5_24}
&&R=\int_0^r \sqrt{-g_{11}}dr=-\int_0^{G_0(r)}  \frac{1}{x^2} \, \sqrt{\frac{\hat g}{g_{00}}} \, \frac{dx}{dG_0} 
\,dG_0\nonumber\\
&&=\int_0^{G_0}\frac{2\mu G^{2(\mu-1)}_0}{(1-G^{2\mu}_0)^2}dG_0<\infty \quad if \quad \mu>1
\end{eqnarray}

It is often required to express metric tensor in conformly-Euclidean system coordinates defined as
$ds^2=\bar g_{00}(y)dt^2-g_c(d\rho^2+\rho^2d\Omega)$. 
For the Jacobson solution this can be done using these formula (Appendix C):
\begin{eqnarray}
\label{f5_25}
&&G_0=(\frac{1-\frac{\mu}{4}y} {1+\frac{\mu}{4}y})^{\frac{1}{\mu}}\quad where\,\, y=1/\rho\nonumber\\
&& x=\frac{y}{(1-\frac{\mu}{4}y)^{1-\frac{1}{\mu}}(1+\frac{\mu}{4}y)^{1+\frac{1}{\mu}}}\\
&&\bar g_{00}(y) \equiv g_{00}(x(y))=G^2_0=(\frac{1-\frac{\mu}{4}y} {1+\frac{\mu}{4}y})^{\frac{2}{\mu}}\nonumber
\end{eqnarray}

As we mentioned above, the Jacobson metric has no singularities. However, when presented in the
 conformly-Euclidean system coordinates it does have singularity at $y=\mu/4$. The reason for that is clearly seen
from the formula x vs. y in eq (\ref{f5_25}, line2). The x(y) transfers $x=\infty$ (r=0) to $y=\mu /4$ ($\rho=4/\mu$).
The singularity of conformly-Euclidean system coordinates is due to our "bad choice" of system coordinates. 
Perhaps, the system coordinates with unity coefficient in front of $dr^2$ ($ds^2=\bar g_{00}(y)dt^2-d\rho^2-g_\Omega(\rho)
d\Omega^2$) is a better choice. As we showed earlier in eq.  (\ref{f5_24}), the function $g_{\Omega} \equiv r^2(\rho)$ is 
regular for all $\rho$ and the parameter $\rho$ is the true distance betwee two points along the radius.
\vskip 3em

One puzzling issue of Einstein's GR, that still has not been resolved, is the definition of the 
energy-momentum tensor of gravitation. It seems logical to identify the tensor $E_{ij}\equiv -(R_{ij}+1/2Rg_{ij})$
(where $R_{ij}$ is the Ricci tensor) as an energy-momentum tensor of the curved space. The Einstein equations
\begin{eqnarray}
&&R_{ij}-1/2Rg_{ij}=T_{ij} \quad or \quad E{ij}+T{ij}=0
\end{eqnarray}
then can be read in this manner: the total energy-momentum tensor of the system (matter and space) is zero.

The difficulty here comes from consideration of the vacuum: $T_{ij}=0$ (no matter) and thus $E_{ij}=0$, which leads
to the unconventional (to say the least) statement that in vacuum gravitation has no energy. 
As we saw in this paper, all the solutions of the pure-aether theory yield Schwarzschild metric, 
which in its turns sets to zero Ricci (and thus $E_{ij}$) tensor. In the Einstein-aether theory, on the other hand,
this problem is resolved. Most of the solutions - and Jacobson's metric (with $G_1=0$) in particular
- yield "almost-Schwarzschild" (up to $x^2$ terms) metric for which $R_{ij}$ (and thus $E_{ij}$) is not zero. 

This seems to be a key factor in resolving the competition between the pure-aether ($\lambda_0=0$) 
and the Einstein-aether ($\lambda_0 \neq 0$) theories in favor of Einstein-aether theory.

\vskip 1em
One more note on a physical nature of space. In the Einstein-aether theory (as in Einstein's GR) space (metric)
is taken as independent physical entity with some energy attached to it. It is expressed in existing "space only"
Lagrangian term ($\lambda_0R$ - the Einstein term). However, in Einstein-aether theory there is another interpretation
of space. We can write the Einstein term in this form:
\begin{eqnarray}
&&L_R =\lambda_0RG_kG^k \equiv \lambda_0R \quad due\,\,to\,\, G_kG^k=1
\end{eqnarray}

In this form the Einstein term ($\lambda_0R$) does not represent space as equal to matter entity,
but rather a part of the aether ($G_i$). The metric, that represents the curved space, is now only auxiliary entity that
ties together all forms of matter including the aether as gravitational matter.

\newpage
\vskip 3em
{\underline {Appendix A}}
\vskip 1em
In this appendix we derive the expression for the Lagrangian of Einstein-Aether theory (see \ref{f5_1})
thru variable $g_{00}$, $\hat g:=-g_{00}g_{11}$ and $G_0$ as a function of $x=1/r$.

In the case of sphirical symmetry the differential of 4-volume $dv$ can be written as:

\begin{eqnarray}
\label{f5_A1}
&& \sqrt{-g} dv=\sqrt{\hat g} \,r^2 dr d\Omega dt=-\sqrt{\hat g}\, \frac{1}{x^4}\, dxd \Omega dt
\end{eqnarray}

The action integral can be written as:
\begin{eqnarray}
\label{f5_A2}
&&S=-\int\nolimits dv L\sqrt{\hat g}=\int\nolimits dtd\Omega S_r \nonumber\\
&&where \quad S_r=-\int\nolimits r^2dr L\sqrt{\hat g}=\int\nolimits dx L(x)\sqrt{\hat g(x)} \quad x=1/r
\end{eqnarray}

Since Lagrangian is only a function of radius r (or x=1/r), to shorten the formula everywhere below in writing
action integral S we will drop the term $d \Omega dt$.

For components of tensor Ricci we have:
\begin{eqnarray}
\label{f5_A3}
&& R_{00}={R^i}_{0i0}={\Gamma}^i_{00,i}-{\Gamma}^i_{0i,0}+{\Gamma}^i_{im}{\Gamma}^m_{00}-{\Gamma}^i_{0m}{\Gamma}^m_{0i}
\nonumber\\
&&={\Gamma}^1_{00,1}+[{\Gamma}^0_{01}+{\Gamma}^1_{11}+2{\Gamma}^2_{21}]{\Gamma}^1_{00}
-[{\Gamma}^1_{00}{\Gamma}^0_{01}+{\Gamma}^0_{01}{\Gamma}^1_{00}]\nonumber\\
&&={\Gamma}^1_{00,1}+[-\frac{g_{00,1}}{2g_{00}}+\frac{g_{11,1}}{2g_{11}}+\frac{g_{22,1}}{g_{22}}]{\Gamma}^1_{00}\nonumber\\
&&=(\frac{ {\Gamma}^1_{00} \sqrt{-g_{11}}g_{22} }{ \sqrt{ g_{00} } })_{,1} \frac{ \sqrt{ g_{00} } }{\sqrt{-g_{11}}g_{22}}\nonumber\\
&&=\frac{g_{00}}{2\hat g r^2} (\frac{ g_{00,1}r^2}{\sqrt{\hat g}})_{,1}=\frac{x^4g_{00}}{2\sqrt{\hat g}}(\frac{g'_{00}}{\sqrt{\hat g}})'
\end{eqnarray}

\begin{eqnarray}
&& R_{11}={R^i}_{1i1}={\Gamma}^i_{11,i}-{\Gamma}^i_{1i,1}+{\Gamma}^i_{im}{\Gamma}^m_{11}-{\Gamma}^i_{1m}{\Gamma}^m_{1i}
\nonumber\\
&&={\Gamma}^1_{11,1}-[{\Gamma}^0_{10,1}+{\Gamma}^1_{11,1}+2{\Gamma}^2_{12,1}]\nonumber\\
&&+[{\Gamma}^0_{10}+{\Gamma}^1_{11}+2{\Gamma}^2_{12}]{\Gamma}^1_{11}
-[{\Gamma}^0_{10}{\Gamma}^0_{10}+{\Gamma}^1_{11}{\Gamma}^1_{11}+2{\Gamma}^2_{12}{\Gamma}^2_{12}] \nonumber\\
&&=-{\Gamma}^0_{10,1}-[{\Gamma}^0_{10}-{\Gamma}^1_{11}+2{\Gamma}^2_{12}]{\Gamma}^0_{01}+ \underline{
\underline{{2\Gamma}^2_{21} 
({\Gamma}^0_{01}+{\Gamma}^1_{11})}}-2\underline{[{\Gamma}^2_{12,1}+{\Gamma}^2_{12}{\Gamma}^2_{12}]}\nonumber
\end{eqnarray}
Double underlined terms can be expressed through new variable $\hat g$ and the single underlined terms cancel each other.

\begin{eqnarray}
\label{f5_A5}
&&=-{\Gamma}^0_{10,1}-[{\Gamma}^0_{10}-{\Gamma}^1_{11}+2{\Gamma}^2_{12}]{\Gamma}^0_{01}+ {2\Gamma}^2_{21} 
(\frac{{\hat g_{,1}}}{2\hat g})\nonumber\\
&&=-(\frac{ {\Gamma}^0_{01} \sqrt{g_{00}}g_{22} }{ \sqrt{ -g_{11} } })_{,1} \frac{ \sqrt{ -g_{11} } }{\sqrt{g_{00}}g_{22}}
+{2\Gamma}^2_{21} (\frac{{\hat g_{,1}}}{2\hat g})\nonumber\\
&&=\frac{g_{11}}{2\hat g r^2} (\frac{ g_{00,1}r^2}{\sqrt{\hat g}})_{,1}+(\frac{{\hat g_{,1}}}{r\hat g})
=\frac{x^4g_{11}}{\sqrt{\hat g}} [(\frac{g'_{00}}{2\sqrt{\hat g}})'+\frac{g_{00}\hat g'}{x{\hat g} \sqrt{\hat g} }]\nonumber\\
&&=\frac{x^4g_{11}}{\sqrt{\hat g}} [(\frac{g'_{00}}{2\sqrt{\hat g}}-\frac{2g_{00}}{x\sqrt{\hat g}})'
+(\frac{g_{00}}{x})' \frac{2}{\sqrt{\hat g}}]
\end{eqnarray}

\begin{eqnarray}
\label{f5_A6}
&& R_{22}={R^i}_{2i2}={\Gamma}^i_{22,i}-{\Gamma}^i_{2i,2}+{\Gamma}^i_{im}{\Gamma}^m_{22}-{\Gamma}^i_{2m}{\Gamma}^m_{2i}
\nonumber\\
&&={\Gamma}^1_{22,1}-{\Gamma}^3_{23,2}+[{\Gamma}^0_{01}+{\Gamma}^1_{11}+2{\Gamma}^2_{21}]{\Gamma}^1_{22}
-[2{\Gamma}^1_{22}{\Gamma}^2_{21}+{\Gamma}^3_{32}{\Gamma}^3_{32}]\nonumber\\
&&={\Gamma}^1_{22,1}+[\frac{g_{00,1}}{2g_{00}}+\frac{g_{11,1}}{2g_{11}}]{\Gamma}^1_{22}\nonumber+
[-{\Gamma}^3_{23,2}-{\Gamma}^3_{32}{\Gamma}^3_{32}]\nonumber\\
&&=\frac{ ({\Gamma}^1_{22} \sqrt{-g_{11}g_{00} })_{,1}} { \sqrt{ -g_{11}g_{00} }}-[(\frac{g_{33,2}}{2g_{33}})_{,2}
+(\frac{g_{33,2}}{2g_{33}})^2]\nonumber\\
&&=\frac{1}{\sqrt{\hat g}} (\frac{ g_{00}r}{\sqrt{\hat g}})_{,1}-
[\frac{(sin^2\theta)_{,\theta} }{2sin^2\theta}]_{,\theta}-[\frac{(sin^2\theta)_{,\theta} }{2sin^2\theta}]^2
=\frac{1}{\sqrt{\hat g}} (\frac{ g_{00}r}{\sqrt{\hat g}})_{,1}+1\nonumber\\
&&=\frac{x^4g_{22}}{\sqrt{\hat g}} \, [(\frac{g_{00}}{x\sqrt{\hat g}})'-\frac{\sqrt{\hat g}}{x^2}]
\end{eqnarray}

Combining expressions (\ref{f5_A3}), (\ref{f5_A5}) and (\ref{f5_A6}) we get this expression for the first term 
($\lambda_0$ term) of the action integral:

\begin{eqnarray}
\label{f5_A7}
&&S_{r {\lambda}_0}=-\int\nolimits \sqrt{\hat g }\, r^2dr \, \lambda_0 R \nonumber\\
&& =\int\nolimits \sqrt{\hat g }\, dx (x^4) \, \lambda_0 [R_{00}g^{00}+R_{11}g^{11}+2R_{22}g^{22}]\nonumber\\
&&=\int\nolimits \, dx \,
 \lambda_0[(\frac{g'_{00}}{\sqrt{\hat g}})'-(\frac{2g_{00}}{x\sqrt{\hat g}})'
+2(\frac{g_{00}}{x})' \frac{1}{\sqrt{\hat g}}-2\frac{\sqrt{\hat g}}{x^2}]
\end{eqnarray}

The first two terms in (\ref{f5_A7}) are full differentials and could be dropped from the expression yielding this:
\begin{eqnarray}
\label{f5_A8}
&&S_{r {\lambda}_0}=\int\nolimits \, dx \,
 \lambda_0[2(\frac{G^2_0-\bar G^2_1}{x})' \frac{1}{\sqrt{\hat g}}-2\frac{\sqrt{\hat g}}{x^2}]
\end{eqnarray}
where per eq. (\ref{f5_4}) we replace $g_{00}$ with $G^2_0-\bar G^2_1$.

\vskip 2em
The $\lambda_1$-term can be straight forward written as:
\begin{eqnarray}
\label{f5_A9}
&&S_{r {\lambda}_1}=\int\nolimits dx\, \lambda_1 [-2\frac{(G_0')^2}{\sqrt{\hat g} }]
\end{eqnarray}

\vskip 2em
For $\lambda_2$-term we get the following expression:
\begin{eqnarray}
&&S_{r {\lambda}_2}=-\int\nolimits r^2dr \, \lambda_2 R_{ij}G^iG^j=\int\nolimits x^4dx \, \lambda_2 R_{ij}G^iG^j\nonumber\\
&&=\int\nolimits d^4x \, \lambda_2 
[R_{00}g^{00}(G_0)^2g^{00}+R_{11}g^{11}(G_1)^2g^{11}]\nonumber\\
&&=\int\nolimits  dx\, \lambda_2 \{(\frac{g'_{00}}{\sqrt{\hat g}})'[(G_0)^2g^{00}+(G_1)^2g^{11}]
+ (\frac{g_{00}\hat g'}{x{\hat g} \sqrt{\hat g} } ) (G_1)^2g_{11} \} \nonumber\\
&&=\int\nolimits  dx\, \lambda_2 \{\underline{ (\frac{g'_{00}}{\sqrt{\hat g}})'}
+ (\frac{g_{00}\hat g'}{x{\hat g} \sqrt{\hat g} } ) (G_1)^2g^{11} \} \nonumber
\end{eqnarray}
In the expression above the underlined term can be integrated out of this expression. In the second term we switch to the variable
$\bar G_1=G_1 \frac{\sqrt{g_{00} } } {\sqrt {-g_{11}} }$ and do a partial integration:

\begin{eqnarray}
\label{f5_A10}
&&S_{r {\lambda}_2}=\int\nolimits  dx\, \lambda_2  (\frac{\hat g'}{x{\hat g} \sqrt{\hat g} } )  (\bar G_1)^2
=\int\nolimits  dx\, (\frac{-2\lambda_2  }{ \sqrt{\hat g} } )  [\frac{ (\bar G_1)^2}{x}]'
\end{eqnarray}

\vskip 2em
The $\lambda_3$ term can be has this form:
\begin{eqnarray}
\label{f5_A11}
&&S_{r\lambda_3}=-\int\nolimits r^2 \sqrt{g} dr \, \lambda_3 (G^k_{;k})^2
=-\int\nolimits r^2 \sqrt{g} dr \, \lambda_3 [\frac{(G_{11}g^{11}\sqrt{g})_{,1}} {\sqrt{g}}]^2\nonumber\\
&&=\int\nolimits dx \,\frac{\lambda_3}{\sqrt{\hat g}} [(\frac{\bar G_1}{x^2})'x^2]^2
\end{eqnarray}

\vskip 2em
Combining the expressions (\ref{f5_A8}) for $S_{r\lambda_0}$, (\ref{f5_A9}) for $S_{r\lambda_1}$, 
(\ref{f5_A10}) for $S_{r\lambda_2}$ and (\ref{f5_A11}) for $S_{r\lambda_3}$,  and introducing $\bar \lambda_2=\lambda_2+\lambda_0$
we get this final expression for the action integral $S_r$:
\begin{eqnarray}
\label{f5_A13}
&&S_r=\\
&&\int\nolimits dx\{\lambda_0[\frac{2}{\sqrt{\hat g}} (\frac{G^2_0}{x})'-\frac{2\sqrt{\hat g}}{x^2}]
-\frac{2\lambda_1(G_0')^2}{\sqrt{\hat g} }-\frac{2\bar \lambda_2}{\sqrt{\hat g}}(\frac {{\bar  G_1}^2}{x})'
+\frac{\lambda_3}{\sqrt{\hat g}}[x^2(\frac{\bar  G_1}{x^2})']^2 \} \nonumber
\end{eqnarray}

\newpage
\vskip 3em
{\underline {Appendix B}}
\vskip 1em
In this appendix we derive the result of Einstein-Aether theory for the case when $\bar G_1=0$.

Variation of Lagrangian eq.(\ref{f5_4}) yields this set of equations:

a) with respect to $\bar G_1$ ($\delta S_r/\delta \bar G_1=0$)
\begin{eqnarray}
\label{f5_B1}
&&\quad \quad 2 \bar \lambda_2\frac{\bar G_1}{x}(\frac{1}{\sqrt{\hat g}})'
+\lambda_3[\frac{x^4}{\sqrt{\hat g}}(\frac{\bar G_1}{x^2})']'\frac{1}{x^2}=0
\end{eqnarray}
which is satisfied if $\bar G_1=0$

b) with respect to $G_0$ ($\delta S_r/\delta G_0=0$)
\begin{eqnarray}
\label{f5_B2}
&&\lambda_1 (\frac{G'_0}{\sqrt{\hat g}})'-\lambda_0\frac{G_0}{x}  (\frac{1}{\sqrt{\hat g}})'=0 \quad or \nonumber\\
&&G''_0=(\frac{G'_0}{2}-\frac{G_0}{2\hat \lambda_1x})\frac{\hat g'}{\hat g}; \,\, where\,\, 
\hat \lambda_1=\frac{\lambda_1}{\lambda_0}
\end{eqnarray}

c) with respect to $\hat g$ ($\delta S_r/\delta \hat g=0$)
\begin{eqnarray}
\label{f5_B3}
&&\hat g=x^2\{-(\frac {G^2_0}{x})'+\frac{\lambda_1}{\lambda_0} (G'_0)^2  
+\frac{\lambda_2}{\lambda_0} (\frac{G^2_1}{x})'-\frac{\lambda_3}{2\lambda_0} [x^2(\frac{\bar G_1}{x^2})']^2 \}\nonumber\\
&&or\,\,with\,\,\bar G_1=0\quad \rightarrow \,\, \hat g=G^2_0-2G_0G'_0x+\hat \lambda_1 (G'_0)^2x^2
\end{eqnarray}

We now introduce a new variable $x=ln(y)$ and write equations (\ref{f5_B2}), (\ref{f5_B3}) as:
\begin{eqnarray}
\label{f5_B4}
&&y=ln(x) \quad G'_0=\dot G_0\frac{1}{x} \quad G''_0=(G_0)\dot {\,} \, \dot {\,} \frac{1}{x^2}-\dot G_0\frac{1}{x^2} \nonumber\\
&&a) \quad \hat g=G^2_0-2G_0\dot G_0+\hat \lambda_1 (\dot G_0)^2\nonumber\\
&&b) \quad (G_0){\dot \,}{\dot \,} -\dot G_0= (\frac{\dot G_0}{2}-\frac{G_0}{2\hat \lambda_1}) \frac{\dot {\hat g}}{\hat g}
\end{eqnarray}

Substituting eq. (\ref{f5_B4}a) in eq. (\ref{f5_B4}b) we will get:
\begin{eqnarray}
\label{f5_B5}
&&[(G_0) \dot{\,}\dot{\,} - \dot G_0][G^2_0-2G_0\dot G_0+\lambda_1 (\dot G_0)^2]=\nonumber\\
&&(\frac {\dot G_0}{2}-\frac{G_0}{2\hat \lambda_1})[2G_0\dot G_0-2(\dot G_0)^2-2G_0 (G_0)\dot{\,}\dot{\,}
+2\lambda_1\dot G_0(G_0)\dot{\,}\dot{\,}]
\end{eqnarray}

And after some algeraic manipulations we will get:
\begin{eqnarray}
\label{f5_B6}
&&(G_0)\dot{\,}\dot{\,}G^2_0=\hat \lambda \dot G^3_0-G_0 \dot G^2_0+G^2_0 \dot G_0
\end{eqnarray}
The equation above has no explicit y-variable and thus can be reduced to the equation of first order by switching
to the new varable $V(G_0)=\dot G_0(y)$:
\begin{eqnarray}
\label{f5_B7}
&&\frac{dV}{dG_0}=\hat \lambda_1(\frac{V}{G_0})^2-\frac{V}{G_0}+1
\end{eqnarray}
And after introducing new variable $\bar V=V/G_0$:
\begin{eqnarray}
\label{f5_B8}
&&\frac{d \bar V}{dG_0}G_0=\hat \lambda_1{\bar V}^2-2{\bar V}+1
\end{eqnarray}
which can be integrated:
\begin{eqnarray}
\label{f5_B8a}
&&\int \nolimits \frac{d \bar V}{\hat \lambda_1{\bar V}^2-2{\bar V}+1}= ln(G_0)+C \quad or \quad \nonumber\\
&&ln(\frac{\bar V-\bar V_1)}{(\bar V-\bar V_2)}=\mu \, ln(G_0)+C
\rightarrow \bar V=\frac{\bar V_1-C\bar V_2G^{\mu}_0}{1-CG^\mu_0}\\
&& ,\quad where\quad \mu=\hat \lambda_1(\bar V_1-\bar V_2); \quad \bar V_{1,2}=\frac{1{\underline +}\sqrt{1-\hat \lambda}}{2\hat \lambda}\nonumber
\end{eqnarray}
In the formula (\ref{f5_B8a}) above C is an integration constant and $\bar V_1$ and $\bar V_2$ are the roots of quadratic
polymon on rhs of eq.(\ref{f5_B8}). Taking into account the expression for $\bar V$ thru $G_0(x)$ we get this equation:
\begin{eqnarray}
\label{f5_B9}
&&\int \nolimits \frac{dG_0(1-CG^{\mu}_0)}{G_0(\bar V_1-C\bar V_2G^{\mu}_0)}=\int \nolimits \frac{dx}{x}
\end{eqnarray}
The constant C above must be chosen as $C=\bar V_1/\bar V_2$ for the reason that lhs of equation above has logafifmic behavior
 at $G_0$ near 1 as rhs at x=0.
\begin{eqnarray}
\label{f5_B10}
&&\int \nolimits \frac{dG_0(1-\frac{\bar V_1}{\bar V_2}G^{\mu}_0)}{G_0\bar V_1(1-G^{\mu}_0)}=\int \nolimits \frac{dx}{x}
\end{eqnarray}
Substituting $U=G^\mu$ the eq. can be written as:
\begin{eqnarray}
\label{f5_B11}
&&\int \nolimits \frac{dU (1-\frac{\bar V_1}{\bar V_2}U)}{U(1-U)}=\mu \bar V_1 ln(x)+C_0 \nonumber\\
&&or \quad \int \nolimits dU[\frac{1}{U}+\frac{(\frac{\bar V_1}{\bar V_2}-1)}{U-1}=ln(C_0X^{V_1\mu}) \nonumber\\
&&or \quad U(U-1)^{\frac{\bar V_1}{\bar V_2}-1}=C_0x^{\mu V_1} \nonumber\\
&&or \quad G^{\frac{1}{\bar V_1}}(G^{\frac{\bar V_1-\bar V_2}{\bar V_1 \bar V_2}}-1)=C_0x
\end{eqnarray}
And substituting in above the values for $\bar V_1$ and $\bar V_2$ thru $\lambda$ we get:
\begin{eqnarray}
\label{f5_B12}
&&G_0(G^{\sqrt{1-\hat \lambda}} - G^{-\sqrt{1-\hat \lambda} })=C_0x
\end{eqnarray}
The sign of $C_0$ should be chosen to satisfy the condition of "attractive gravity"
 - $G_0$ decreases as x increases from zero on.

\newpage
\vskip 3em
{\underline {Appendix C}}
\vskip 1em
The goal of this appendix is to derive the expression for the Jacobson metric in conformly-Euclidean
coordinates.
The Jacobson metric is a spherically symmetrical metric given by this expression - see eq. (\ref{f5_10}).
\begin{eqnarray}
\label{f5_C1}
&&g_{00} \equiv g_{tt}=G^2_0; \quad C_0x=G(G^{-\mu}-G^{\mu})\quad x=\frac{1}{r} \nonumber\\
&&g_{11}\equiv g_{rr}=-\frac{\hat g}{g_{00}} \quad \hat g=\frac{4{\mu}^2G^2_0}{[(1-\mu)G^{-\mu}-(1+\mu)G^{\mu}]^2} \\
&&C_0=const. \quad \mu=\sqrt{1-\lambda} \quad \lambda<0\nonumber
\end{eqnarray}
The transition to conformly-Euclidean coordinates (r$\rightarrow \rho$) is done according to this equation:
\begin{eqnarray}
\label{f5_C2}
&&ds^2=\bar g_{00}dt^2-(\frac{r}{\rho})^2dl^2 \quad l=Euc lidean \,\,length \nonumber\\
&&where \quad \bar g_{00}(\rho)=g_{00}(r(\rho))\nonumber\\
&&and \,\, r(\rho) \,\, satisfies \,\, \sqrt{-g_{11}} \,\, \frac {dr}{d \rho}=\frac{r}{\rho}
\end{eqnarray}

The eq. \ref{f5_C2} can be first written in x and y coordinates (x=1/r; y=1/$\rho$) and then in $G_0$, y coordinates:
\begin{eqnarray}
\label{f5_C3}
&&\sqrt{ \frac{\hat g}{g_{00}} } \,\, \frac {dx}{x}=\frac{dy}{y}\nonumber\\
&&\sqrt{ \frac{\hat g}{g_{00}} } \,\, \frac {dx}{xdG_0}dG_0=\frac{dy}{y}
\end{eqnarray}

Substituting expressions eq.(\ref{f5_C1}) in to eq.(\ref{f5_C3}) we will get:
\begin{eqnarray}
\label{f5_C3a}
&& \frac{2\mu}{G_0[G^{\mu}_0-G^{-\mu}_0]}dG_0=\frac{dy}{y}\nonumber\\
&&2 \frac{du}{u^2-1}=\frac{dy}{y} \quad where \quad u=G^{\mu}_0<1
\end{eqnarray}

or after integration:
\begin{eqnarray}
\label{f5_C4}
&& ln(\frac{1-u}{1+u})=ln(y)+C \nonumber\\
&& or\quad y=\frac{4}{\mu}( \frac{1-G^{\mu}_0}{1+G^{\mu}_0})  \quad G_0=
(\frac{1-\frac{\mu}{4}y} {1+\frac{\mu}{4}y})^{\frac{1}{\mu}}\nonumber\\
&&and \quad g_{00}=G^2_0=(\frac{1-\frac{\mu}{4}y} {1+\frac{\mu}{4}y})^{\frac{2}{\mu}}
\end{eqnarray}

where the constant C is taken as $C=\frac{\mu}{4}$ so $g_{00}$ at small y (large $\rho$) has approximation
$g_{00}=1-y$. From here we can find the transformation coordinates $x \rightarrow y$:
\begin{eqnarray}
\label{f5_C5}
&& x=\frac{y}{(1-\frac{\mu}{4}y)^{1-\frac{1}{\mu}}(1+\frac{\mu}{4}y)^{1+\frac{1}{\mu}}}
\end{eqnarray}

IF $\mu=1$ ( the case of GR) the expression (\ref{f5_C5}) becomes:
\begin{eqnarray}
\label{f5_C6}
&& x=y \frac{1}{(1+\frac{1}{4}y)^2} \quad or \quad r=\rho (1+\frac{1}{4\rho})^2
\end{eqnarray}
which is a well know expression from the theory of GR \cite{i5}.

\end{document}